\documentstyle[graphicx]{article}

\begin{document}
\title{Distinguishability, contrast and complementarity in
multimode two-particle interferences}
\author{Pedro Sancho \\ GPV de Valladolid \\ Centro Zonal en
Castilla y Le\'on \\ Ori\'on 1, 47014, Valladolid, Spain}
\maketitle
\begin{abstract}
Multimode two-particle systems show interference effects in one-
particle detections when both particles have common modes. We
explore the possibility of extending the usual concepts of
distinguishability and visibility to these types of systems.
Distinguishability will refer now to the balance between common and
different modes of a two-particle system, instead of the standard
definition concerning available alternatives for a one-particle
system. On the other hand, the usual concept of visibility is not
suitable for our problem and must be replaced with that of contrast,
measuring the ratio of detection probabilities with and without
interference effects. Finally, we show that for the type of states
considered in the paper there is a complementarity relation between
distinguishability and contrast for two-boson states. In contrast
there is not a two-fermion counterpart.
\end{abstract}
\vspace{7mm}
\newpage
\section{Introduction}
Distinguishability, visibility and their complementarity relation
have become fundamental concepts in the understanding of
interferometric properties of quantum systems \cite{GY,JSV}. In
closely related developments, the demonstration of the
complementarity between the distinguishability of the different
paths available to a particle and the visibility of the interference
patterns obtained in the detectors was a cornerstone in the
formulation of the which-way experiments \cite{E,BK}. The importance
of the complementarity relations is confirmed by their experimental
verification (see, for instance \cite{DNR}) and by their robustness
against imperfections in the experimental arrangements (see, for
instance, \cite{S} for a complementarity relation in the presence of
imperfections).

Recently, in a different context, the potential impact of multimode
systems on two-photon interferometry has been signaled. Manipulating
the composition of the state of the photons, one can modify the
detection patterns, a result that could have interesting
applications in quantum imaging and other related subjects. For
instance, in \cite{bra} the Hong-Ou- Mandel experiment \cite{HOM}
was realized with photons in multimode states.

In this paper we want to analyse if the concepts of visibility and
distinguishability can be extended to multimode systems. Many
different types of interference arrangements could be considered.
However, in order to simplify the analysis and to highlight the main
concepts involved, we shall concentrate on a particular case, the
detection at fixed positions of one of the particles of a
two-particle system. This way, we focus on the interference effects
associated with the existence of common modes to the two particles.
We shall show that for this particular arrangement the correct
notion of distinguishability must refer to the measure of the
distinguishability between the two particles composing the complete
system, which is related to the balance between common and different
modes of the two particles. On the other hand, as we shall also see
later, the notion of visibility usual in standard interferometry (or
the extension to continuous variables \cite{Ralph}) is not suitable
for our problem. Instead, we introduce the concept of contrast,
which compares the detection probabilities when the interference
effects are, or are not, taken into account. These are the natural
extensions of the concepts of distinguishability and visibility for
our type of arrangement. We must then explore the existence of
complementarity relations between distinguishability and contrast.
At this point we shall find that the relations are not general. To
be concrete, for the type of states considered in this paper the
relations exist for two-boson systems. In contrast they do not exist
for two-fermion systems.

The main novelty of our approach is that we do not restrict our
analysis, as it is usual in these types of problems, to one-particle
systems. Instead, we consider two-particle systems (note that in
\cite{JSV} a complementarity relation was established for
two-particle systems, but the variables involved were one- and
two-particle visibilities instead of visibility and
distinguishability). Moreover, our notion of distinguishability
refers to modes instead of available alternatives or paths.

The plan of the paper is as follows. In section 2 we present the
arrangement and evaluate the detection probabilities for multimode
states of the two-particle system. As an illustration, we consider
explicitly the case of Gaussian distributions. Sections 3 and 4
deal, respectively, with the definitions of distinguishability and
contrast. In section 5 we analyse the existence of complementarity
relations. Finally, in the conclusions we discuss the results of the
paper.

\section{Detection probabilities}

First of all we describe the arrangement (see Fig. 1). It is
composed of two independent sources emitting particles of the same
type in multimode states. Interferometric experiments with two
photons coming from different sources have already been realized
\cite{PM,RMT}, but we will consider massive particles. At a fixed
position in the intersection of the directions of propagation of the
two particles we place a small size detector (ideally point-like)
which can measure the arrival of particles. We restrict our
considerations to the cases when only one of the particles is
detected at that position, disregarding the events in which both
particles are detected. Experimentally, the distinction between one-
and two-particle detection events can be achieved placing other
secondary detectors after the main one, covering all the possible
directions and registering the times of detection. If the interval
between two repetitions of the experiment is large enough as
compared to the typical time interval between events in the main
detector and the secondary ones in every single repetition of the
experiment we can discern between both types of detection events.

\begin{figure}[p]
\center
\includegraphics[width=15cm, height=12cm]{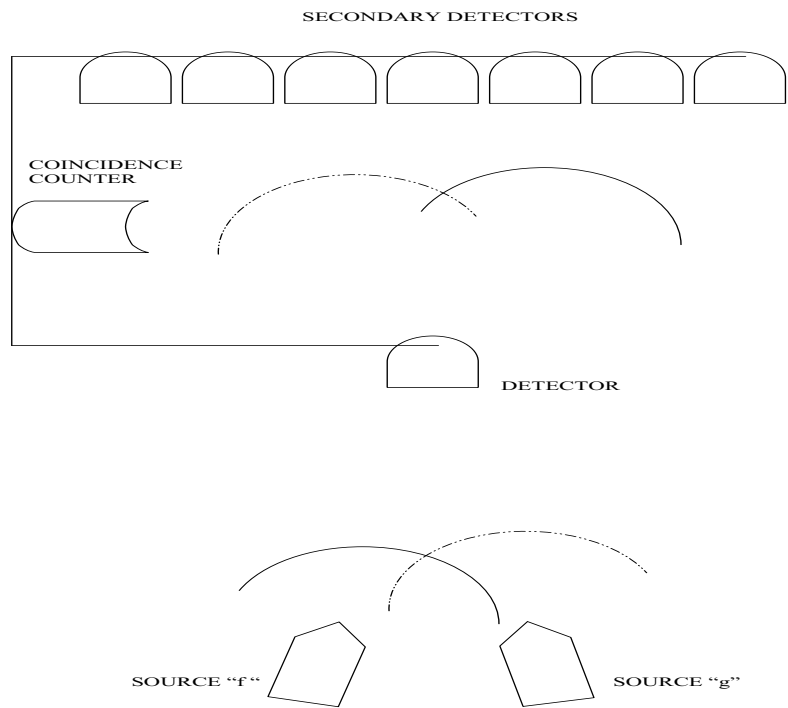}
\caption{Arrangement for the detection of the particles. The set
of secondary detectors covers all the directions where the
detection probability is not negligible. The main detector and
the secondary ones are connected to a coincidence counter to
distinguish between events at the main detector triggered by one
of two particles. The continuous and discontinuous circumference
arcs represent, respectively, the particles "f" and "g".}
\end{figure}
The second quantization formalism is a powerful technique for the
description of multiparticle systems. In this framework, the
particles are mathematically represented by the Schr\"{o}ndiger
field operator. If we restrict to the stationary problem it is
given by
\begin{equation}
\hat{ \psi} ({\bf r}) = \int d^3 {\bf q} \psi _{\bf q} ({\bf r})
\hat{a}
({\bf q})
\end{equation}
where the $\psi _{\bf q} ({\bf r}) $ form a complete basis of
stationary wavefunctions labelled by the continuum index ${\bf q}$.
On the other hand, $\hat{a} ({\bf q}) $ is the annihilation operator
of the state ${\bf q}$. A common choice for the basis is that of
plane waves parametrized by the momentum ${\bf p}$:
\begin{equation}
\hat{ \psi} ({\bf r})= \frac{1}{(2 \pi \hbar)^{3/2}} \int d^3
{\bf
p} \exp{\left( \frac{i{\bf p}.{\bf r}}{\hbar } \right) } \hat{a}
({\bf p})
\label{eq:DDD}
\end{equation}
The initial state in Fock's space after the emission of the
particles is (both particles are emitted by independent sources):
\begin{equation}
|I>=\int d^3 {\bf p} \int d^3 {\bf q} \; f({\bf q}) g({\bf p})
\hat{a} ^+ ({\bf q}) \hat{a}^+ ({\bf p}) |0>
\end{equation}
In this expression $|0>$ is the vacuum state and $f$ and $g$ are
two functions of ${\bf q}$ and ${\bf p}$. When $f({\bf q}) \neq
\delta ({\bf q}-{\bf q_o})$ and $g({\bf p}) \neq \delta ({\bf p}-
{\bf p_o})$ with ${\bf q_o}$ and ${\bf p_o}$ arbitrary we deal
with multimode states.

The parameters ${\bf p}$ and ${\bf q}$ refer to the same index used
to parametrize the Schr\"{o}ndiger field operator. Every value of
the parameters represents a different mode. With this choice every
mode is a component of the same basis of wavefunctions used to write
down the field operator  As we only consider the stationary problem
the modes refer to stationary wave functions, i. e., solutions of
the one-particle stationary Schr\"{o}dinger's equation.

In order to simplify
the presentation we assume that, $f$ and $g$ are real and non-
negative functions. In this way they can be identified as the
squared roots of the mode distributions, i. e., $f^2({\bf q})$
($g^2({\bf p})$) gives the weight of the mode ${\bf q}$ (${\bf
p}$) in the superposition of modes of the particle. The more
general case of $f({\bf q})$ being a complex function
(representing now $|f({\bf q})|^2$ the weight in the
distribution) can be handled in a similar way, but the mathematical
presentation would be more lengthy.

Note that in the above expression we have not explicitly included
spin indexes. This is equivalent to assume that both particles in
the same spin state. We use this simplification in order to clarify
the analysis.

The probability of detection of only one particle at point ${\bf
r}$ can be evaluated in the usual way \cite{Gal}
\begin{equation}
P({\bf r})=\frac{<I| \hat{\psi}^+({\bf r}) \hat{\psi}({\bf r})
|I>}{<I|I>}
\label{eq:cu}
\end{equation}
In this paper we shall refer to this probability as the detection
probability. It gives the probability of detecting at point ${\bf
r}$ one particle in a single repetition of the experiment. It is not
to be confused with the detection efficiency, usually used in
quantum optics or matter waves interferometry. The detection
probability is closely related to the measured intensity, which is
proportional to the number of detections in a large number of
repetitions of the experiment. From the experimental point of view
the detection probability is obtained from the measured intensity.

A better understanding of the physical meaning of the above formula
can be obtained by comparison with similar expressions in quantum
optics. An expression proportional to $<\hat{E} ^- \hat{E} ^+ >$,
with $\hat{E}=\hat{E} ^+ + \hat{E} ^-$ the electric field operator,
gives the rate of detection of photons by a phototube \cite{Loudon}.
In particular, for single-mode beams the expectation value,
$<\hat{E} ^- \hat{E} ^+ >$, is proportional to the mean number of
photons in the light beam. Consequently, the detection rate is
proportional to the mean number of photons. Physically, this
proportionality can be understood because "the phototube measures
the intensity of a light beam by counting the rate at which atomic
ionizations take place, which is in turn proportional.....to the
beam intensity" \cite{Loudon}. When multimode beams are considered
$<\hat{E} ^- \hat{E} ^+ >$ is no longer proportional to the mean
number of photons, but can be interpreted as the observable beam
intensity \cite{Loudon}. The detection rate is proportional to this
observable intensity. In the case of massive particles the
discussion follows similar lines. For particles in single-mode
states the numerator of the r. h. s. of equation (\ref{eq:cu}) is
proportional to the mean number of particles in the state $|I>$.
Therefore, assuming a connexion between the probability of detection
and the mean number of particles similar to that discussed by Loudon
between the detection rate and the mean number of photons we arrive
at the usual interpretation of equation (\ref{eq:cu}), which is
extended to multimode states.

The calculation of the detection probability can be easily
carried out using the (anti)commutation relations:
\begin{equation}
[\hat{a}({\bf p}),\hat{a}^+({\bf q})]_{\mp}=\delta ^3({\bf p}-
{\bf q})
\end{equation}
In all the expressions with a double sign the upper and lower
ones will refer, respectively, to bosons and fermions.

Note that in the last equation the spin indexes have not been
included. This is a consequence of the previous assumption that both
particles are in the same spin state. Then the general relation
$[\hat{a}({\bf p},s),\hat{a}^+({\bf q}, s')]_{\mp}=\delta ^3({\bf
p}- {\bf q}) \delta _{ss'}$ transforms into the above equation.

First, we evaluate the denominator of Eq. (\ref{eq:cu}) (as
usual, the vacuum is normalized to unity $<0|0>=1$):
\begin{equation}
<I|I>=\int d^3 {\bf p} \int d^3 {\bf q} (f({\bf q}) f({\bf p})
g({\bf q}) g({\bf p}) \pm f^2({\bf q}) g^2({\bf p}))
\label{eq:se}
\end{equation}
We assume the functions $f$ and $g$ to be normalized to unity
\begin{equation}
\int f^2({\bf p}) d^3 {\bf p} =1= \int g^2({\bf p}) d^3 {\bf p}
\label{eq:enueva}
\end{equation}
This is the usual normalization for one-particle states, where the
spatial normalization $\int |\psi ({\bf r})|^2 d^3 {\bf r} =1 $,
with $\psi ({\bf r})= \int f({\bf p}) \psi _{{\bf p}} ({\bf r}) d^3
{\bf p}$ gives, via the orthonormality conditions, $\int \psi _{{\bf
p}}^* ({\bf r}) \psi _{{\bf q}} ({\bf r})  d^3 {\bf r} = \delta
^3({\bf p}-{\bf q})$, equation (\ref{eq:enueva}).

With this choice equation (\ref{eq:se}) becomes
\begin{equation}
<I|I>=\pm 1 + \int d^3 {\bf p} \int d^3 {\bf q} f({\bf q}) f({\bf
p}) g({\bf q}) g({\bf p})
\end{equation}
This scalar product cannot be factored into the product of two
one-particle scalar products. Moreover, when $f=g$ we have $<I|I>=0$
for fermions. This corresponds to the physically forbidden situation
of two identical fermions. It is simple to show that the scalar
product is zero or negative for fermions (property 1 in appendix A).
The situation is similar to that found in the quantization of the
free electromagnetic field \cite{IZ}. However, the physically
measurable magnitudes are the detection probabilities, which as we
shall see later, are positive definite and, consequently, property 1
does not represent a problem at all (this scalar product would be
positive by introducing $|I> \rightarrow i|I>$ for fermions, but
then the notation for bosons and fermions would be different).

Evaluating the numerator of equation (\ref{eq:cu}) results in the
detection probability expression
\begin{eqnarray}
P({\bf r})=\alpha _{fg} P_{fg}({\bf r})+\alpha _{fg} P_{fg}^*
({\bf
r}) \pm \alpha _{gg} P_{ff}({\bf r}) \pm \nonumber  \\
\alpha _{ff} P_{gg}({\bf r})= 2\alpha _{fg} Re P_{fg}({\bf r})
\pm \alpha _{gg} P_{ff}({\bf r}) \pm \alpha _{ff} P_{gg}({\bf r})
\label{eq:si}
\end{eqnarray}
where $Re(.)$ refers to the real part of the complex expression
$(.)$,
\begin{equation}
\alpha _{fg} = \frac{\beta _{fg}}{<I|I>}
\end{equation}
with
\begin{equation}
\beta _{fg} = \int d^3 {\bf q} f({\bf q}) g({\bf q}) \; ; \;
\beta _{ff}=\beta _{gg}=1
\end{equation}
and
\begin{equation}
P_{fg}({\bf r}) = \int d^3 {\bf q} \int d^3 {\bf p} f({\bf q})
g({\bf p}) \psi _{{\bf q}}^* ({\bf r}) \psi _{{\bf p}} ({\bf r})
\end{equation}
The form of equation (\ref{eq:si}) is clearly that of an
interference phenomenon. $P_{ff}$ and $P_{gg}$ are the detection
probabilities one obtains when only particles of one of the two
sources are emitted and detected. These probabilities are weighted
by the coefficients $\alpha _{gg}$ and $\alpha _{ff}$. On the other
hand, $2\alpha _{gf} Re P_{gf}$ is the interference term. It
contains the contributions of the product of the wavefunctions of
the two particles.

We note that the terms $P_{ff}$ and $P_{gg}$ would also be produced
by a mixture of one-particle states, but the interference term
clearly shows that we are dealing with a two-particle pure state.

When the particles have no common modes ($f\cap g=\o$) the
interference term vanishes, property that agrees with the well-
known origin of this type of interferences: the detector is unable
to distinguish if a mode common to both particles has its origin in
one or other of the sources. In quantum mechanics two
indistinguishable alternatives give rise to interference patterns.

Note that in the case of fermions with equal mode distributions
($f=g$) the expression for the probability is undefined because
both, the numerator and denominator of equation (\ref{eq:cu}), are
null. Physically this corresponds to the impossibility of measuring
any property of a system of two identical fermions, since the two
particles cannot be prepared in that two-particle state. We discuss
this point to more extent in the example presented below and in
appendix B.

We also remark that the detection probability is non-negative for
fermions in spite of the fact that the denominator of equation
(\ref{eq:cu}) is negative or zero. This property that follows
directly from the definition of the detection probability, can
easily be checked from equation (\ref{eq:si}) which can be rewritten
for fermions as $|\alpha _{ff}|P_{gg}+|\alpha _{gg}|P_{ff} +2 \alpha
_{gf} Re P_{gf}$. Then using equations (\ref{eq:aci}) and
(\ref{eq:ase}) of the appendix (presented there for other purpose)
we can easily see that this expression is non-negative.

We end this section presenting an example where all these
expressions can be evaluated analytically. We assume both
distribution functions to be Gaussians with equal spread:
\begin{equation}
f({\bf p})=N \exp{(-({\bf p}-{\bf f_o})^2/Q^2)};g({\bf p})= N
\exp{(-({\bf p}-{\bf g_o})^2/Q^2)}
\end{equation}
with $N= (2/\pi Q^2)^{3/4}$ and $Q$ being constant factors, and
${\bf f_o}$ and ${\bf g_o}$ two constant vectors.

Using repeatedly the well-known formula $\int _{\infty}^{\infty}
exp(\mu x^2+\eta x)dx=(\pi /-\mu)^{1/2} exp(-\eta ^2 /4 \mu)$ for
$Re(\mu) \leq 0$ we easily obtain:
\begin{equation}
<I|I>= \pm 1 + exp ( {-({\bf f_o}-{\bf
g_o})^2/Q^2} )
\end{equation}
\begin{equation}
\beta_{fg}= \exp{(-({\bf f_o}-{\bf g_o})^2/2
Q^2)}
\end{equation}
and, finally, using as basis of the wave functions plane waves (see
equation (\ref{eq:DDD}))
\begin{eqnarray}
P({\bf r})=\frac{Q^3}{\sqrt{8} \hbar ^3}
\exp{(-Q^2{\bf r}^2/2\hbar ^2)} \times \nonumber \\
\left( \frac{ \pm 1 + \exp{(-({\bf f_o}-{\bf g_o})^2/2 Q^2)} cos
( ({\bf f_o}-{\bf g_o}).{\bf r})/\hbar ) }{ \pm 1 + \exp{(-({\bf
f_o}-{\bf g_o})^2/Q^2)} } \right)
\label{eq:ds}
\end{eqnarray}
We can easily check with this explicit model two characteristics
found previously, the scalar product is negative or zero for
fermions and the detection probability for fermions tends to an
undefined expression of the type $0/0$ when ${\bf f_o} \rightarrow
{\bf g_o}$. We show in appendix B that this indetermination cannot
be removed by the application of L'H\^{o}pital's rule

Equation (\ref{eq:ds}) shows clearly the existence of
interference effects. The interference term is that with the
cosine function. For an experiment with detectors placed at
different positions we would have an oscillatory dependence on
the position variable. However, we are only considering an
experiment with the position of the detector fixed. Now, we can
observe an oscillatory dependence in the detections at the fixed
point if we vary ${\bf f_o}-{\bf g_o}$, i. e., for different
compositions of the multimode states.

Note that this example can be described as a single mode situation
if we consider a Gaussian modes basis (every particle is then in a
single mode state). However, the results are similar for other
multimode states that cannot be reduced to single mode states in
particular modes basis.

Let us consider in the next sections the possibility of extending
the concepts of distinguishability and visibility and their
complementarity relations to this type of interference phenomena.
\section{Distinguishability}
Let us briefly review the definition of distinguishability in
standard interferometry, for instance in the usual beam splitter
experiment. The particle is prepared in a two-dimensional Hilbert
space: $c_1 |1> + c_2 |2>$, representing $|1>$ and $|2>$ the two
alternatives in the beam splitter, the two available pathways.
Varying $c_1$ and $c_2$ we obtain different intensities at the
detector placed at a fixed position, corresponding to an
interference phenomenon. The distinguishability of the two
alternatives is $||c_1|^2-|c_2|^2|$ \cite{JSV}.

In our arrangement we are concerned with modes instead of paths. The
distinguishability must now refer to the differences between the
mode distributions of both particles. Moreover, we distinguish
between two particles instead of the usual approach, which considers
alternatives of a one-particle system. A definition of
distinguishability suitable for our experiment must meet three
conditions:

1) If there are no common modes the distinguishability must reach
its maximum value: both particles are completely distinguishable.

2) When the two particles are prepared in the same state (equal
distribution of modes) the distinguishability must be minimal: both
particles are identical an indistinguishable.

3) Following the usual convention the values of the
distinguishability must be in the interval $[0,1]$.

We propose the following definition of distinguishability, which
fulfills the three above criteria ($\beta _{fg}$ positive and $\beta
_{fg} \leq 1$ (equation (\ref{eq:aci})) in the appendix A):
\begin{equation}
D=1-\frac{2\int f({\bf p}) g({\bf p}) d^3 {\bf p}}{\int f^2({\bf
p}) d^3 {\bf p}+\int g^2({\bf p}) d^3 {\bf p}} = 1- \beta _{fg}
\end{equation}
In the example of the Gaussian distribution this expression
becomes
\begin{equation}
D=1- \exp{(-({\bf f_o} - {\bf g_o})^2/2Q^2)}
\end{equation}
that has value $0$ for $f=g$. For this type of distributions we do
not have the case of no common modes because these distributions are
never strictly zero. However, in the limit $2Q^2/({\bf f_o} - {\bf
g_o})^2 \rightarrow 0$, in which the number of common modes tends to
zero, we have asymptotically $D \rightarrow 1$.
\section{Contrast}
In standard interferometry the measure of the visibility of the
interference patterns is given by $(I_{max}-I_{min})/(I_{max} +
I_{min})$, with $I_{max}$ and $I_{min}$ being the maximum and
minimum values of the intensity (or probability of detection
events). In the case of interference by a beam splitter the
variation of the intensity is due to changes of values of $c_1$ and
$c_2$ (see the first paragraph in the previous section). In
diffraction granting experiments different intensity is obtained at
each point after the granting.

In our case, by analogy, the visibility would be given by
$V_p=(P_{max}-P_{min})/(P_{max}+P_{min})$, with $P_{max}$ and
$P_{min}$ being the maximum and minimum values of the detection
probability. However, this is not a satisfactory definition for our
problem The aim of the paper would be a complementarity relation
between $V_p$ and $D$. The distinguishability, as deduced in the
previous section, is a function of $f$ and $g$: for every pair of
distributions $f$ and $g$ we obtain a value of $D$. Therefore, an
appropriate definition of visibility should assign a value of
visibility to every pair of distributions. This is not the case for
$V_p$: once $f$ and $g$ are given, as the position of the detector
is fixed, the detection probability has a unique value. There are no
maximum or minimum values of the detection probability, but a single
value. We could compare the different values of the detection
probability obtained varying $f$ and $g$ to determine $P_{max}$ and
$P_{min}$. This way, however, we cannot assign a value of $V_p$ to
every pair, $f$ and $g$, of distributions.

Other measure of visibility valid for continuous variables has been
presented in the quantum optics literature, \cite{Ralph}. When broad
band beams are considered their states are characterized by their
distribution of power between phase and amplitude fluctuations. In
the case of two-photon systems this power distribution refers to the
intensity fluctuation correlations. These measurements are done via
a homodyne detection system instead of photons counters. As in our
arrangement we focus on the detection by "particle counters" the way
of measuring in \cite{Ralph} shows that it is not a measure of
visibility appropriate for our problem.

We shall introduce the new measure, showing in this section that it
is physically acceptable. Later, in the next section, we shall see
that the new measure gives complementarity relations in some cases,
justifying its introduction {\em at posteriori}.

In order to see how to do it we shall consider another interference
experiment, the usual two-slit experiment with only one particle and
with a fixed position for the detector. The source emits one
particle in every repetition of the experiment, which travels to a
screen with two slits. A detector is placed at a fixed position
after the screen. If $\phi _A$ and $\phi _B$ denote the
wavefunctions of the particle passing respectively through slit A or
B (both normalized), the complete wavefunction of the particle is
$(\phi _A + \phi _B)/ \sqrt{2} $, and the detection probability at
the fixed position ${\bf r}$ of the detector becomes $\tilde{P}
({\bf r})= \frac{1}{2}|\phi _A ({\bf r}) |^2 + \frac{1}{2} | \phi _B
({\bf r}) |^2 +Re(\phi _A^* ({\bf r}) \phi _B ({\bf r}) )$. As all
the parameters of the problem are constant (the position  of the
detector and the wavefunction of the particle are fixed for all the
repetitions of the experiment) the usual concept of visibility
cannot be applied because we do not have variation of the
probability detection and there are no maximum and minimum values.
We have two contributions to the detection probability $\tilde{P} _o
({\bf r})= \frac{1}{2}|\phi _A ({\bf r}) |^2 + \frac{1}{2} | \phi _B
({\bf r}) |^2$ and $\tilde{P} _I ({\bf r})= Re(\phi _A^* ({\bf r})
\phi _B ({\bf r}) )$. The first one corresponds to the detection
probability in the absence of interference effects, i. e., the
detection probability obtained when in every repetition of the
experiment we close one of the slits avoiding one of the two
alternatives for the particle and, consequently, also the
interference effects. The interference effects are associated with
$\tilde{P} _I$. A natural measure of the interference effects is
given by comparing the probability detections in the presence and
absence of the interference effects, i. e., by $\tilde{P}({\bf
r})/\tilde{P}_o({\bf r}) $ (valid for points where $\tilde{P}_o
({\bf r}) $ is not null). When there are no interference effects the
ratio is 1. When the ratio is different from 1 we have interference
effects whose strength is measured by the separation of the ratio
with respect to 1. Other possible measure would be of the form $|
\tilde{P} _I ({\bf r})| / \tilde{P}_o ({\bf r}) $ (the absolute
value of $\tilde{P} _I$ must be used in order to maintain the
non-negative character of the measure). However, this definition
does not distinguish between contributions of the interference
effects that increase or decrease the detection probability.
Consequently we disregard it.

Now we return to our original problem. Following the same
arguments considered in the two-slit problem, the measure of the
importance of the interference contribution to the probability
detection must be based on the ratio between both types of
probability detections. We propose the following definition which
we shall name as contrast
\begin{equation}
C=\frac{P}{P_o}=1+\frac{2\alpha _{fg} Re(P_{fg})}{|\alpha _{gg}|
P_{ff} +|\alpha _{ff}| P_{gg}} =1 \pm \frac{2\beta _{fg}
Re(P_{fg})}{ P_{ff} + P_{gg}}
\end{equation}
where $P$ is given by equation (\ref{eq:si}) and $P_o=|\alpha _{gg}|
P_{ff} +|\alpha _{ff}| P_{gg} $. Note the absolute values for the
$\alpha $ in the last expression; in this way we assure the positive
definite character of $P_o$ and $C$ for fermions. We have also used
the property $\alpha _{fg} =\pm |\alpha _{fg}|$ in the last step.

We analyse now some of the properties of the new definition. First,
it is positive or zero as follows directly from the definition. It
is defined at all the points where $P_o \neq 0$. We do not care
about the singular points $P_o=0$. The values of the contrast are
within the interval $[0,2]$. It is very simple to demonstrate the
last statement taking into account that $\tilde{C}$ (defined by $C=1
\pm \tilde{C}$) obeys the relation $|\tilde{C}| \leq 1$ (see
property 2 of Appendix A). When $2\beta _{fg} Re(P_{fg})=0$ there
are no interference effects and $C=1$. This condition is obtained in
particular when there are no common modes for the two particles.
This behaviour is similar for bosons and fermions. However, for
other situations the behaviour is completely different. Let us
consider the case $f=g$ (for fermions $f \rightarrow g$ because the
limit $f=g$ is forbidden). For bosons $f=g$ leads to $C=2$. The
detection probability doubles that without interference effects. On
the other hand, for fermions with $f \rightarrow g$, $ C \rightarrow
0$ reflecting the property that, contrarily to bosons, the detection
probability tends to vanish. Because $Re(P_{fg})$ can be positive or
negative we can have values $C>1$ or $C<1$ for both, bosons and
fermions.

At first sight $|\tilde{C} |=|C-1|$ would also be a good definition
of "visibility". However, there are two properties of $|\tilde{C}|$
that invalidate it as an acceptable candidate to replace the
visibility. The first one is that it cannot distinguish between
contributions of the interference term that enhance or diminish the
detection probability. The argument follows closely that presented
for $| \tilde{P} _I ({\bf r})| / \tilde{P}_o ({\bf r}) $ in the case
of the one-particle two-slit experiment. Depending on the fact of
$Re(P_{fg})$ being positive or negative, the contributions to the
detection probability have opposite sign. However, when we use
$|\tilde{C}|$ we lost the information about the sign of the
contribution of the interference term. We do not know if it tends to
increase or to decrease the detection probability. Now, we consider
the second property. One would expect that for large detection
probabilities, i. e., for large increments of the detection
probability due to the contribution of the interference term, the
value of the variable measuring the contrast would increase. This is
so for bosons, $|\tilde{C}| \rightarrow 1$ when $f \rightarrow g$
and the detection probability reaches its maximum value. However,
for fermions when $f \rightarrow g$, $|\tilde{C}|$ tends to $1$, the
maximum value of $|\tilde{C}|$; in spite of the fact that the
detection probability tends to zero, its minimum value.

Finally, let us briefly discuss how to evaluate the contrast ratio.
$P({\bf r})$ is directly obtained from the experimental detections
values. $P_o({\bf r})$ can be evaluated combining the experimental
determination of $P_{ff}$ and $P_{gg}$, and the knowledge (obtained
in the preparation) of $\alpha _{ff}$ and $\alpha _{gg}$. The method
is to carry out two different experiments with one-particle states.
In the first one only one of the sources emits particles in the
state $|f>=\int d^3 {\bf q} f({\bf q}) \hat{a}^+ ({\bf q})|0>$. The
probability detection is $P_{ff}$. In the same way, in the second
experiment using particles in the state $|g>$ we obtain
experimentally $P_{gg}$. Now, with $\alpha _{ff}$ and $\alpha
_{gg}$, which can be calculated from the preparation of the state,
we obtain $P_o({\bf r})$, and we can evaluate the contrast.
Equivalently, the contrast could be determined using $Re(P_{fg})$,
which can be obtained from equation (\ref{eq:si}) using a method
similar to that described above.

Based on the above considerations we conclude that contrast is a
good measure of the quality of the interference effects considered
in this paper. This is so because of three main reasons: (i)It gives
a quantitative estimation of the contribution of the interference
effects to the detection probability. (ii) As we shall see later, it
allows for complementariry relations with distinguishability. (iii)
It can be easily evaluated from the experimental data of detection
and the knowledge of the initial state ($f$ and $g$ distributions).
Contrast is a good measure for our arrangement, and also for the
two-slit experiment discussed at the beginning of this section, i.
e. , for experiments where the detection is restricted to a fixed
position. If it is a good measure for other type of interference
experiments should be analysed in every arrangement.
\section{Complementarity relations}
In standard interferometry there are complementarity relations
between the distinguishability of the different paths available
to the particle and the visibility of the interference patterns.
We now look for the existence of complementarity relations
between the distinguishability of the two particles and the
contrast of the probability detection in our arrangement.

To analyse this problem we shall rest on the following inequality:
\begin{equation}
\frac{2 \beta _{fg} Re(P_{fg})}{ P_{gg}+ P_{ff}} \leq \beta _{fg}
\label{eq:vu}
\end{equation}
This inequality is equivalent to $2 Re(P_{fg})/( P_{gg}+ P_{ff})
\leq 1 $ (equation (\ref{eq:ase}) in the appendix A). Note that we
can assume $Re(P_{fg}) \geq 0$ because if $Re(P_{fg}) < 0$, equation
(\ref{eq:vu}) automatically holds.

We analyse first the case of bosons. We have $C  \leq 1+ \beta
_{fg}$, and
\begin{equation}
D+ C  \leq 2
\label{eq:vcu}
\end{equation}

This is a typical complementarity relation. When the
distinguishability increases the contrast diminishes; a large
distinguishability corresponds to a small number of common modes and
small interference effects $P$ being close to $P_o$. In the limit of
$D=1$ we have $C=1$. On the other hand, when $D$ is small we have a
large number of common modes and large interference effects. In
particular, in the limit $D=0$ we have $C=2$, the maximum value of
the contrast.

The equality sign in the above expression is valid when equation
(\ref{eq:vu}) is an equality: $2Re(P_{fg})=P_{ff}+P_{gg}$, relation
valid for $f=g$, i. e., when both particles have the same
distribution and for some particular values of the $P$ when $f\neq
g$.

We note that visibility and distinguishability enter in the
standard complementarity inequality in a quadratic way, instead
of the linear one found in the previous inequality. However, this
is not a fundamental difference because our variables can be
redefined as $D_* =D^{1/2}$ and $C_* =C^{1/2}$, becoming the
above inequality $D_*^2 + C_*^2 \leq 2$.

We consider now the case of fermions. We obtain
\begin{equation}
D+C \geq 2 ( 1-\beta _{fg})
\end{equation}
Now, we do not have a complementarity relation. Physically, the
non existence of a
complementarity relation can be easily understood. Both,
distinguishability and contrast show the same behavior (increase
or decrease simultaneously) when their compositions ($f$ and $g$)
vary, in contrast with the opposite behavior of bosons. For
instance, when the number of common modes increases both, $C$
and $D$, diminish, the second one because unlike the case of
bosons the interference contribution tends to cancel that of
$P_o$.

We note that using $|\tilde{C}|$ as measure of the contrast one
has complementarity relations for both, bosons and fermions.
However, as discussed in the previous section, $|\tilde{C}|$ is
not an acceptable measure of the contrast.

We conclude that there is a complementarity relation in the case
of bosons. In contrast, such relation does not exist for bosons.

\section{Conclusions}

We have analysed in this paper the interference effects associated
with the existence of common modes in the detection of one of the
members of a multimode two-particle system. For simplicity we have
restricted our considerations to real and non- negative functions
$f$ and $g$, and states with equal spin for both particles. This
particular case illustrates the most fundamental properties of the
problem. The extension to the general problem where complex
distributions $f$ and $g$ and other states with different spin
values for the particles (which allow for other
symmetric-antisymmetric combinations of the spin and spatial degrees
of freedom of the particles) follows along similar lines but would
be much more lengthy and will be considered in future work. The
arrangement discussed in this paper is in the same line with the
experiments considered in \cite{bra,HOM,PM,RMT}. The main novelties
of our arrangement rest on the facts that the interference effects
can be observed studying only the detection properties of one of the
two particles and that it is directed to massive particles. It is
important to note the relation with the Hanbury Brown-Twiss (HBT)
experiment \cite{HBT}, and related intensity correlation
arrangements. In our case we compare the events at the detector and
the secondary detectors only in order to avoid double detection
events in the main detector, not as in HBT to measure their
correlations. Moreover, the comparison is not carried out as in HBT
at two points, but at a point (where the main detector is placed)
and at a large number of points (where the secondary detector are
placed covering all the possible directions for the particles). The
interference effects are not associated with the
indistinguishability of the two particles, but with the existence of
common modes (even in the case that both particles are not identical
because their spectral composition is not equal). The experimental
realization of this type of arrangement would be an interesting
further step in two-particle interferometry with independent
sources.

We have studied the possibility of extending the usual notions of
distinguishability and visibility to our arrangement. The definition
of distinguishability here proposed is different from the usual one
in standard interferometry in two main aspects. First, it refers to
the distinguishability of the two particles of the system instead of
the usual distinguishability of a one- particle system. Second, it
is based on the existence of common modes, not in the possibility of
following different alternatives or available paths the particle.

We have also shown that the usual notion of visibility is not
suitable for our problem. We have replaced it by the concept of
contrast, based on the comparison of the detection probabilities
with and without interference effects. The analytical form of the
contrast as a function of $\beta _{fg}$ is different for bosons and
fermions. We have also explored the existence of complementarity
relations between distinguishability and contrast. They are obtained
in the case of bosons (provided that the distribution functions obey
equation (\ref{eq:enueva})). These complementarity relations are an
example of complementarity for the distinguishability in
two-particle systems. Note that in the case of two-particle systems
a complementarity relation has been demonstrated \cite{JSV}. However
the complementary variables are one- and two-particle visibilities,
not distinguishability. In contrast, there is no complementarity
relation for fermions. This is a new manifestation of the different
behaviour of both types of particles. Of course, this result does
not mean the impossibility of finding other measures of "visibility"
suitable for two-fermion systems in arrangements as the one here
proposed or in other experiments previously carried out
\cite{Kiesel}. These measures would give rise to complementarity
relations with the distinguishability. This possibility is an open
problem that should be addressed in future work.

Finally, we shall consider the feasibility of experimentally
carrying out the ideal arrangement presented in the paper. We focus
on two points, the preparation of the beams and the data acquisition
time. Photon beams can be prepared in multimode states (see, for
instance, \cite{bra}). In the case of massive particles, trapping
techniques can be used to prepare matter waves in multimode states.
For instance, a thermal atomic cloud confined in a trap of width $L$
is described by a state superposition of the eigenfunctions of the
trap. When the trap is opened adequately, the resulting beam can
propagate in the desired direction. Each eigenfunction can be
approximated by a wave packet with mean momentum proportional to
$(n+1)/L$ ($n=0, 1,..$), and momentum spread proportional to $1/L$
\cite{suec}. Using this type of techniques it seems possible, in
principle, to prepare beams in multimode states with specific
distributions as those described in this paper. On the other hand,
we have the problem of reaching permissible data acquisition times.
These limits have been considered by several authors (see, for
instance, \cite{japo}), showing that for standard time windows of
the coincidence counter the data acquisition time is within
permissible limits. An important simplification of the detection
scheme would be achieved if it would be experimentally demonstrated
that the rate of double detections is negligible when compared to
that of single events. Then we could eliminate the secondary
detectors and the coincidence counter, shortening in a notorious way
the data acquisition time. Once you have prepared the two particles
in multimode and equal spin states, the experiment reduces to a
simple detection and counting arrangement. In similar arrangements,
but with only one source emitting in each repetition of the
experiment, we can determine $P_{ff}$ and $P_{gg}$. Collecting all
these data we can evaluate the contrast. On the other hand, the
distinguishability is deduced from the preparation of the particles.
Comparing both values we can test the relation between
distinguishability and contrast. We conclude that making available
sources of multimode states, the arrangement considered in previous
sections could be experimentally realized.

{\bf Acknowledgments}

This work has been partially supported by the DGICyT of the
Spanish Ministry of Education and Science under Contract No.
REN2000-1621 CLI. I am grateful to Joan Alcaide for helping in
the preparation of the manuscript.

{\bf Appendix A: Two mathematical properties}

In this Appendix we demonstrate the two following properties:

PROPERTY 1. For fermions $<I|I> \leq 0$.

We present a general demonstration of this property independent of
the normalization of equation (\ref{eq:enueva}). For fermions we
have equation (\ref{eq:se}) with the negative sign. We introduce the
notation $N_f =\int d^3{\bf q} f^2 ({\bf q})$, $N_g =\int d^3{\bf q}
g^2 ({\bf q})$, $F({\bf q})=f({\bf q})/ \sqrt{N_f}$ and $G({\bf
q})=g({\bf q})/ \sqrt{N_g}$.

Using the inequality $(F({\bf q})- G({\bf q}))^2 \geq 0$ we have
\begin{equation}
F({\bf q}) G({\bf q}) \leq \frac{1}{2} F^2({\bf q})+ \frac{1}{2}
G^2({\bf q})
\end{equation}
Using these expressions we obtain
\begin{equation}
\int F({\bf q}) G({\bf q}) d^3 {\bf q} \leq 1
\end{equation}
and, finally
\begin{equation}
\int d^3 {\bf q} \int d^3 {\bf p} f({\bf q}) g({\bf q}) f({\bf
p}) g({\bf p})  \leq N_f N_g
\end{equation}
which is equivalent to
\begin{equation}
<I|I> \leq 0
\end{equation}

PROPERTY 2. $|\tilde{C}| \leq 1$.

First, using as in property 1 that $(f({\bf q})- g({\bf q}))^2 \geq
0$ we obtain (taking into account equation (\ref{eq:enueva}))
\begin{equation}
\beta _{fg} \leq 1
\label{eq:aci}
\end{equation}
On the other hand, we have
\begin{equation}
| 2 Re(P _{fg})| \leq P _{ff} + P _{gg}
\label{eq:ase}
\end{equation}
The last expression can be easily verified using the complex
functions
\begin{equation}
\phi _+ = \int d^3 {\bf q} f({\bf q}) \psi _{{\bf q}} + \int d^3
{\bf q} g({\bf q}) \psi _{{\bf q}}
\end{equation}
and
\begin{equation}
\phi _- = \int d^3 {\bf q} f({\bf q}) \psi _{{\bf q}} - \int d^3
{\bf q} g({\bf q}) \psi _{{\bf q}}
\end{equation}
Then as $\phi _+  \phi _+^* = P_{ff} + P_{gg}+ 2 Re(P_{fg}) \geq 0$
and $\phi _- \phi _-^* = P_{ff} + P_{gg}- 2 Re(P_{fg}) \geq 0$ we
obtain equation (\ref{eq:ase}) noting that $P_{ff}\geq 0$ and
$P_{gg}\geq 0$.

Finally using equations (\ref{eq:aci}) and (\ref{eq:ase}) we obtain
for $\tilde{C}=2\beta _{fg} Re(P_{fg})/(P_{ff}+P_{gg})$
\begin{equation}
|\tilde{C}| \leq 1
\end{equation}

{\bf Appendix B: Application of L'H\^{o}pital's rule}

It is usual to resort to L'H\^{o}pital's rule to elucidate the
correct limit of expressions whose numerator and denominator tend
simultaneously to $0$. This is the case of equation (\ref{eq:ds})
for fermions when ${\bf f_o} \rightarrow {\bf g_o}$. With the
notation $W_1$, $W _2$ and $W _3$ for the three components of the
vector ${\bf f_o}-{\bf g_o}$, equation (\ref{eq:ds}) can be
expressed as $P({\bf W}) \sim F({\bf W}) = P_N({\bf W})/P_D({\bf
W})$ with $P_N({\bf W})=-1+exp(-{\bf W}^2/2Q^2)cos({\bf W}.{\bf
r}/\hbar )$ and $P_D({\bf W})=-1+exp(-{\bf W}^2/Q^2)$. In order that
the limit exists when ${\bf W} \rightarrow {\bf 0}$ it must exist
for any succession of ${\bf W}$ that tends to ${\bf 0}$. Let us
consider, for instance, the succession $W _1^* \rightarrow 0$, $W
_2^* =0$ and $W _3^*=0$. This way we reduce the problem to an
one-dimensional one and we can use L'H\^{o}pital's rule in the
standard way. As $lim _{W _1^* \rightarrow 0} P_N({\bf W ^*}) = lim
_{W _1^* \rightarrow 0} P_D({\bf W ^*}) =0$, L'H\^{o}pital's rule
states that if $ lim _{W _1^* \rightarrow 0} (P_N'({\bf W
^*})/P_D'({\bf W ^*}))$ exists, it is the limit of $F({\bf W ^*})$
for ${\bf W ^*}\rightarrow {\bf 0}$. The evaluation is simple
\begin{eqnarray}
P_N'({\bf W ^*})=\left( \frac{dP_N({\bf W ^*})}{dW _1^*} \right)=
-\frac{W _1^*}{Q^2} exp(-(W _1^*)^2 /2 Q^2) \times \\
cos(W _1^* x_1/\hbar)-\frac{x_1}{\hbar} exp(-(W _1^*)^2 /2 Q^2) sin(W _1^* x_1/\hbar) \nonumber
\end{eqnarray}
and
\begin{equation}
P_D'({\bf W ^*})=\frac{-2 W _1^*}{Q^2} exp(-(W _1^*)^2 /Q^2)
\end{equation}
In the limit $W _1^* \rightarrow 0$ both expressions tend to zero and we obtain again an indeterminate expression for their ratio. We apply again L'H\^{o}pital's rule and, with an obvious notation we obtain
\begin{eqnarray}
P_N''({\bf W ^*})=exp(-(W _1^*)^2 /2 Q^2) cos(W _1^* x_1/\hbar) \left( -\frac{1}{Q^2}-\frac{x_1^2}{\hbar ^2}+\frac{(W _1^*)^2}{Q^4} \right) \nonumber \\
+\frac{2 W _1^* x_1}{\hbar Q^2} exp(-(W _1^*)^2 /2 Q^2) sin(W _1^* x_1/\hbar)
\end{eqnarray}
\begin{equation}
P_D''({\bf W ^*})=\frac{-2}{Q^2} \left( 1- \frac{2 (W _1^*)^2}{Q^2} \right) exp(-(W _1^*)^2 /Q^2)
\end{equation}
The limit of $F({\bf W ^*})$ when $W _1^* \rightarrow 0$ is
\begin{equation}
lim _{W _1^* \rightarrow 0} F({\bf W^*}) = \frac{1}{2} \left(
1+\frac{x_1^2 Q^2}{\hbar ^2 } \right)
\end{equation}
Now we can consider a different succession $\tilde{W} _1=\tilde{W}
_3 =0$ and $\tilde{W} _2 \rightarrow 0$, reducing the problem again
to a one-dimensional one, now in the axis "2". Following the same
steps the limit is now
\begin{equation}
lim _{\tilde{W} _2 \rightarrow 0} F({\bf \tilde{W}}) = \frac{1}{2}
\left( 1+\frac{x_2^2 Q^2}{\hbar ^2 } \right)
\end{equation}
Clearly the limits for both successions are different (and so on for
any other direction). Consequently, the limit when ${\bf W}
\rightarrow {\bf 0}$ is not defined in a unique way: we have a
different limit for every direction. $P({\bf W})$ remains undefined
at ${\bf W}={\bf 0}$. Physically, this indetermination in the limit
${\bf W} \rightarrow {\bf 0}$ reflects the impossibility of
obtaining definite quantum predictions for two-fermion systems in
equal states, a two-particle state forbidden in quantum theory.

\end{document}